\documentclass[prd,twocolumn,preprintnumbers,showpacs,amsmath,amssymb,nofootinbib]{revtex4}
\usepackage{citesort,latexsym,ifthen,graphics,color,hyperref,epsfig}

\newcommand{\nothing}{}

\newcommand{\ie}{{i.e.}}
\newcommand{\cf}{{cf.}}
\newcommand{\eg}{{e.g.}}

\newcommand{\etc}{{etc.}}

\newcommand{\wrt}{with respect to}

\newcommand{\lhs}{left-hand side}

\newcommand{\rhs}{right-hand side}

\newcommand{\be}{\begin{equation}}
\newcommand{\ee}{\end{equation}}
\newcommand{\bea}{\begin{eqnarray}}
\newcommand{\eea}{\end{eqnarray}}
\newcommand{\beas}{\begin{eqnarray*}}
\newcommand{\eeas}{\end{eqnarray*}}

\newcommand{\bear}{\begin{array}{l}}
\newcommand{\eear}{\end{array}}

\newcommand{\bcf}{\begin{center}\begin{figure}}
\newcommand{\ecf}{\end{figure}\end{center}}

\newcommand{\bct}{\begin{center}\begin{table}}
\newcommand{\ect}{\end{table}\end{center}}

\newcommand{\ds}{\displaystyle}

\def\eq#1{(\ref{eq:#1})}

\def\eqs#1#2{(\ref{eq:#1}) and~(\ref{eq:#2})}

\def\fig#1{figure~\ref{fig:#1}}

\newcommand{\Int}[1]{\int \!\! d^D \! #1 \,}
\newcommand{\FourInt}[1]{\int \!\! d^4 \! #1 \,}

\newcommand{\der}[2]{\ensuremath{\frac{d #1}{d #2}}}

\newcommand{\fder}[2]{\ensuremath{\frac{\delta #1}{\delta #2}}}
\newcommand{\Or}{\mathcal{O}}

\newcommand{\order}[1]{\Or ( #1 )}
\newcommand{\hf}{\frac{1}{2}}

\newcommand{\measure}[1]{\mathcal{D} #1 \, }

\def\dd{\dot{\Delta}}

\def\hS{\hat{S}}
\def\e#1{\,{\rm e}^{\displaystyle #1}}
\def\one{\hbox{1\kern-.8mm l}}

\newcommand{\DV}{\mathcal{D}}
\newcommand{\OPI}{\overline{\mathcal{D}}}

\newcommand{\itp}{\Delta^{-1}}
\newcommand{\Ker}[1]{\{#1\}}
\newcommand{\DiagDot}{\scriptstyle \bullet}
\newcommand{\DummyKernel}{\ensuremath{\stackrel{\bullet}{\mbox{\rule{1cm}{.2mm}}}}}

\newcommand{\flow}{\Lambda \partial_\Lambda}

\newcommand{\totalflow}{\Lambda \der{}{\Lambda}}

\newcommand{\dec}[3][0]{\ensuremath{\left[ #2 \hspace{#1em} \right]^{#3}}}
\newcommand{\norm}{\ensuremath{\Upsilon}}

\newcommand{\psl}{p \hspace{-.51em}/ }
\newcommand{\ksl}{k \hspace{-.51em}/ }

\newcounter{Diagrams}
\Alph{Diagrams}
\newtheorem{Diag}{}[Diagrams]

\newlength{\VertexWidth}

\newlength{\LabLength}

\newcommand{\arXiv}[2]{arXiv:{#1} [#2]}

\newcommand{\hepth}[1]{hep-th/#1}
\newcommand{\hepph}[1]{hep-ph/#1}
\newcommand{\heplat}[1]{hep-lat/#1}

\begin{document}

\preprint{DIAS-STP-08-01}

\title{A Resummable $\beta$-Function for Massless QED}

\author{Oliver J.~Rosten}
\email{orosten@stp.dias.ie}
\affiliation{Dublin Institute for Advanced Studies, 10 Burlington Road, Dublin 4, Ireland}

\begin{abstract}
	Within the set of schemes defined by generalized, manifestly gauge invariant exact renormalization 
	groups
	for QED, it is argued that the $\beta$-function in the four dimensional
	massless theory cannot possess any nonperturbative power corrections. 
	Consequently, the perturbative expression for the
	$\beta$-function must be resummable. This argument cannot be extended to flows of 
	the other couplings or to the anomalous
	dimension of the fermions and so perturbation theory does not define a unique trajectory in the
	critical surface of the Gaussian fixed point. Thus, resummability of the $\beta$-function is not
	inconsistent with the expectation 
	that a non-trivial fixed point does not exist.
\end{abstract}


\maketitle

%\section{Introduction}

The resummability\footnote{Throughout this paper we have in mind Borel resummability,
though our conclusions should not depend on this choice.}, or otherwise, of the perturbative series for the $\beta$-functions and
anomalous dimension(s)  in some quantum field theory (QFT) is intimately related to the nonperturbative question of renormalizability.
 This is beautifully explained in~\cite{TRM-Elements} (see also~\cite{B+B}), and we here recall the main points. The formalism best suited to understanding such issues is the Exact Renormalization Group (ERG), which is essentially the continuous version of Wilson's RG~\cite{Wilson,WH}. A fundamental ingredient of this approach is the implementation of a momentum cutoff, such that all modes above the cutoff scale are regularized. For the following discussion, we consider two cutoff scales. First, there is the bare scale, $\Lambda_0$, which provides an overall cutoff to the theory. As we shall see, for nonperturbatively renormalizable theories, this scale is an artificial construction, and it is misleading to identify the action at this scale as a boundary condition that can be chosen, arbitrarily. (The same is not true for nonrenormalizable theories.) We now integrate out degrees of freedom between the bare scale and a lower, `effective' scale, $\Lambda$. As we perform this procedure, the bare action evolves into the Wilsonian effective action, $S_\Lambda$, in such a way that the partition function stays the same. The Wilsonian effective action can be thought of as parametrizing the interactions relevant to the effective scale. The ERG equation states how the Wilsonian effective action changes with the effective scale.

One of the most important uses of the ERG equation is to find QFTs which are nonperturbatively renormalizable, in other words theories for which $\Lambda_0$ can be send to infinity (this is called taking the continuum limit). Scale independent renormalizable theories  follow immediately from fixed points of the ERG equation. To see this,  suppose that we rescale all dimensionful quantities to dimensionless ones, by dividing by $\Lambda$ raised to the appropriate scaling dimension. Now, fixed points of the ERG can be immediately identified with renormalizable theories: as a consequence of our rescalings, independence of $\Lambda$ implies independence of all scales; independence of all scales trivially implies independence of $\Lambda_0$, and so obviously we can send $\Lambda_0$ to infinity!

Scale dependent renormalizable theories can be constructed by considering flows out of any of the fixed points, along the associated relevant / marginally relevant directions. The Wilsonian effective actions lying on these `Renormalized Trajectories' (RTs)~\cite{Wilson} are self-similar, meaning that all dependence on $\Lambda$ appears only through the renormalized relevant / marginally relevant couplings and anomalous dimension(s).\footnote{Any masses are included in our definition of couplings.} Self-similarity implies renormalizability, since there is no explicit dependence on $\Lambda/\Lambda_0$. Note that, along an RT, the theory is completely specified by the choice of fixed point, and the integration constants or `rates' associated with the relevant / marginally relevant directions. In the limit $\Lambda \rightarrow \infty$, the theory sinks back into the appropriate fixed point. Thus, if we wish to consider the action at some arbitrarily high `bare' scale, we must \emph{compute} it using the flow equation, given our aforementioned choices. 
Indeed, the `bare action' in this context is the perfect action~\cite{perfect} in the vicinity of the ultraviolet (UV) fixed point. This is in contrast to a nonrenormalizable trajectory, where we can simply chose some bare action, and use it as the boundary condition for the flow.

One of the benefits of viewing renormalization in this way is that, along RTs, we can compute directly in terms of renormalized quantities, without any mention of the bare scale or the bare action. 
To do this, we employ renormalization conditions for the relevant and marginally relevant couplings and the anomalous dimension(s)
directly at the effective scale, $\Lambda$. So, if a non-trivial RT were to exist in QED (we are not claiming that one does in four dimensions, where the gauge coupling is marginally \emph{irrelevant}, but one does in three dimensions) then we would define the gauge coupling---which we denote by $g$ and not $e$ to avoid later confusion--- simply by writing the gauge kinetic term as
\[
	\frac{1}{g^2(\Lambda)} \Int{x} F_{\mu\nu} F_{\mu \nu},
\]
at all scales. Note that we have scaled the coupling out of the gauge field. In the manifestly gauge invariant approach that we will later adopt, this will have the pleasant effect of guaranteeing that the gauge field does not suffer from field strength renormalization~\cite{qed}. Throughout this paper  we work in Euclidean space, so there is no distinction between upper and lower indices.

Let us now consider a massless theory, about which it is supposed that all we know is that its ERG trajectory lies in the critical surface of some fixed point. Since this trajectory is flowing \emph{into} a fixed point, and we have not specified whether or not the trajectory happens to have emanated from some other fixed point in the UV, we do not know, a priori, whether the theory is renormalizable or not. To be concrete, we will suppose that this infrared fixed point is the Gaussian one, that this fixed point possesses a single marginally irrelevant coupling, $g$, and that there is a single anomalous dimension
(just as there is in our manifestly gauge invariant approach to QED in four dimensions).

Let us now do perturbation theory in the vicinity of the Gaussian fixed point, within the critical surface. For reasons that will become apparent, we will attempt to write the action in self-similar form. Consequently, our renormalization conditions involve conditions for only the coupling, $g$, and the anomalous dimension, $\gamma$, specified at the scale $\Lambda$. We have assumed, temporarily, that no reference to the bare scale / bare action is necessary. Computing the full perturbative solution to the theory, we find that everything can be written in renormalized terms~\cite{qed}, defining 
an apparently unique, self-similar trajectory in the critical surface of the Gaussian fixed point. Were it really the case that this trajectory were both self-similar and unique, then this would suggest the existence of a UV fixed point, out of which an RT can be constructed that flows into the Gaussian fixed point. However, as emphasised in~\cite{TRM-Elements}, this picture is generally false. In the specific case of scalar field theory in four dimensions, the
perturbative series for the flows of the $n$-point couplings (the $\beta$-functions) of the theory
are not resummable, and so do not unambiguously define functions. The reason for this is UV renormalons% 
\footnote{Throughout this paper, we will loosely refer to a renormalon as \emph{any} singularity of the Borel transform, rather than using the strict definition~\cite{BenekeReview}, which defines
renormalons as those singularities related to large or small loop momentum behaviour.}
(see~\cite{BenekeReview} for a review of renormalons)%
: perturbation theory by itself is not well defined, but must be supplemented by exponentially small terms
which, in QED, take the form
\be
	 \frac{\Lambda}{\Lambda_0} \sim \e{-1/ 2 \beta_1 g^2(\Lambda)} + \ldots,
\label{eq:suppressed}
\ee
where $\beta_1$ is the one-loop $\beta$-function and the ellipsis denotes higher order corrections
(`the' $\beta$-function refers to the flow of $g$.). The \lhs\ of this expression makes it immediately clear that self-similarity of our trajectory is violated: there is explicit dependence on $\Lambda_0$. 
We can always write such power corrections in terms of $g$, as we have done on the \rhs, but the
prefactor will, of course, depend on $\Lambda_0$.
Within perturbation theory, we can blithely take the limit $\Lambda_0 \rightarrow \infty$, but at some point have to face up to the fact that this procedure is not well defined, if we hope to draw reliable nonperturbative conclusions. 

%(Note that for nonperturbatively renormalizable theories, where the bare scale has been sent to infinity, terms which are both exponentially small in the coupling can nevertheless exist, as they do in QCD.)

In light of this discussion, the main result of this paper is rather unexpected: it will be demonstrated, in massless QED in four dimensions that, given a particular definition of the coupling,
the perturbative series for the $\beta$-function \emph{cannot be supplemented by terms of type~\eq{suppressed} and its generalizations}. 
Since our ERG equation is perfectly well defined, and since we can choose a perfectly well defined
boundary condition (bare action) we deduce that the perturbative $\beta$-function must be resummable. Nevertheless, our earlier conclusions are still intact, since our argument does not 
apply to the other couplings of the theory or to the anomalous dimension of the fermions.
Consequently, perturbation theory still does not specify a unique, self-similar trajectory within the critical surface of the Gaussian fixed point, and so there is no suggestion that a UV fixed point exists. 

Note, though, that matters could be much more interesting in the Wess-Zumino model. 
First, we note that all couplings belonging to the superpotential are protected from flowing by
the nonrenormalization theorem. Secondly, it seems as though we can apply the arguments
of this paper to show that the perturbative series for the anomalous dimension is resummable.
(By scaling the field strength renormalization out of the two-point vertex, we can induce a flow
of the superpotential and so relate the $\beta$-function of the three-point coupling to the
anomalous dimension. It looks like the arguments applying to the $\beta$-function in this
paper go through similarly in the Wess-Zumino model.) 
Finally, following~\cite{NonRenorm},
we can demonstrate that the flow of the dressed, exact $n$-point vertices can be written in
terms of the apparently resummable anomalous dimension [for an example of a dressed two-point
function, see~\eq{D}, below]. Moreover, the relationship between the dressed vertices
and the Wilsonian effective action vertices can be straightforwardly inverted~\cite{NonRenorm}.
This suggests the existence of a (well defined) self-similar trajectory in the critical surface, which would
indicate the presence of a UV fixed point. Work on this is underway~\cite{WIP}.

That terms of the type~\eq{suppressed} are precluded comes about as follows. The key is to express the $\beta$-function as a ratio of two other functions [see~\eq{beta-partial}, below]. Now, there is no reason
to suppose that each of these two functions cannot, separately,
possess contributions of the form~\eq{suppressed}. However, for reasons that we will precisely spell out later, any such contributions must exactly cancel each other.

The definition of our QED coupling is defined through our choice of ERG. In this paper, we use the framework of generalized ERGs~\cite{TRM+JL,mgierg1}, which can be used to furnish a manifestly gauge invariant formulation of QED~\cite{qed} and even non-Abelian gauge theories~\cite{ym1,aprop,mgierg1,qcd,conf}. The essence of this approach is as follows. As stated already, a necessary ingredient of the ERG equation is that the partition function is invariant under the flow. Consequently, given some set of fields, $\varphi$, we can define the family of ERGs to which Polchinski's equation~\cite{pol} belongs according to~\cite{TRM+JL}
\be
\label{eq:blocked}
-\flow \e{-S[\varphi]} =  \int_x \fder{}{\varphi(x)} \left(\Psi_x[\varphi] \e{-S[\varphi]}\right),
\ee
where the $\Lambda$ derivative
is performed at constant $\varphi$,
any Lorentz indices \etc\ have been suppressed
and we have written $S_\Lambda$ as just $S$.
It is the total derivative on
the \rhs\ ensures that the 
partition function $Z = \int \measure{\varphi} \! \e{-S}$
is invariant under the flow.

The functional, $\Psi$, parametrizes a
general Kadanoff blocking~\cite{Kadanoff} in the 
continuum, for which we take the
following form~\cite{mgierg1}:
\be
\label{eq:Psi}
	\Psi_x = \hf \Ker{\dd^{\varphi \varphi}(x,y)} \fder{\Sigma}{\varphi(y)},
\ee
where it is understood that we sum over all the elements
of the set of fields $\varphi$. Whilst we will leave the
blocking procedure largely unspecified, there are certain
general requirements that must be satisfied. Crucially,
blocking must take place only over a localized patch, ensuring
that each infinitesimal RG step is free of IR divergences.

We now describe each of the components on the \rhs\ of~\eq{Psi}. 
First, there are the ERG kernels,
$\dd^{\varphi \varphi}$, which are generally
different for each of the elements of $\varphi$. Each kernel
incorporates a cutoff function which provides UV
regularization. The notation $\Ker{\dd}$ denotes a covariantization
of the kernel which may be necessary, depending on the
symmetries of the theory. Indeed, it is
apparent from~\eqs{blocked}{Psi} that the kernel
essentially ties together two functional derivatives
at points $x$ and
$y$; in gauge theory, we can covariantize this statement
by using \eg\ straight Wilson lines between these
two points. In practice, we leave any necessary covariantization
unspecified, demanding only that it satisfies general
requirements~\cite{ym1,aprop,mgierg1}. The remaining ingredient
in~\eq{Psi} is $\Sigma \equiv S - 2\hS$,
where $\hS$
is the seed action~\cite{aprop,mgierg1,mgierg2,scalar2,qed}.
Whereas we solve the flow equation for the Wilsonian effective action,
the seed action serves as an input and, given our choice~\eq{Psi}
and  a choice of cutoff function(s),
parametrizes the remaining freedom in how modes are
integrated out along the flow. 

The constraint that $\Psi$
corresponds to a local blocking transformation translates into
the requirement that the seed action leads to convergent momentum
integrals and that the seed action and (covariantized) cutoff functions 
have all orders derivative expansions. In turn, this guarantees
that the Wilsonian effective action vertices have a derivative expansion, also,
this being  a property that we will exploit, later.

The final point to make about~\eq{blocked} is a subtle one: 
it might be necessary to include some unphysical fields in the
set $\varphi$, in order to properly implement a UV cutoff. 
Indeed, this is precisely the case in the 
manifestly gauge invariant ERG
formulation of QED that we employ, where
covariantization of the cutoff functions is
not sufficient to completely regularize the
theory: it is necessary to include Pauli-Villars (PV) partners for
the fermions. (This is due to the well known result that covariant higher derivatives
fail to regularize a set of one loop divergences.) Consequently, the field content 
for our manifestly gauge invariant ERG for QED comprises
the gauge field, $A_\mu$, a fermion field, $\psi$, and an 
unphysical commuting spinor field, $\chi$, which is given a mass at the
effective scale (it is obviously trivial to generalize
to extra flavours). 
To be completely clear: when we loosely refer to QED, we strictly mean regularized theories of an Abelian vector field, coupled to fermions, whose effective action in the vicinity of the Gaussian fixed point is that of QED,
to excellent approximation.

The precise details of the set-up can be found in~\cite{qed},
but we will not need them here. Rather, for our purposes, we need only consider the
flow equation for the various \emph{vertex coefficient functions} (\ie\ all
fields and symmetry factors having been stripped off), which has a generic
diagrammatic form, largely independent of the details of the set-up and 
even the precise field theory being considered~\cite{scalar2,qed,qcd}.

Given the aforementioned
field content, we substitute~\eq{Psi} into~\eq{blocked}, perform the 
$\Lambda$-derivative on the \lhs\ and identify terms with the same numbers
of fields~\cite{ym1,aprop,thesis,mgierg1,qed,qcd}. Before doing this,
we scale the coupling out of the covariant derivative, for reasons mentioned
earlier. The rescaling causes $S \rightarrow S / g^2$ and, in contrast to
some previous works~\cite{aprop,mgierg1,qcd,univ}, we choose to similarly
redefine the seed action. Thus, defining $\Sigma_g \equiv g^2(S - 2\hS)$, the
diagrammatic flow equation for the vertex coefficient functions is shown in
\fig{Flow}~\cite{qed}.
\bcf[h]
	\[
	\left(
		-\totalflow + \sum_{\phi \epsilon \{f\}} \!\! \gamma^{(\phi)}
	\right)
	\dec{
		\ensuremath{\begin{array}{c}\begin{picture}(0,0)%
\includegraphics{pstex/Vertex-S.pstex}%
\end{picture}%
\setlength{\unitlength}{3947sp}%
\begingroup\makeatletter\ifx\SetFigFont\undefined%
\gdef\SetFigFont#1#2#3#4#5{%
  \reset@font\fontsize{#1}{#2pt}%
  \fontfamily{#3}\fontseries{#4}\fontshape{#5}%
  \selectfont}%
\fi\endgroup%
\begin{picture}(341,318)(2180,-963)
\put(2291,-859){\makebox(0,0)[lb]{\smash{{\SetFigFont{11}{13.2}{\rmdefault}{\mddefault}{\updefault}{\color[rgb]{0,0,0}$S$}%
}}}}
\end{picture}%
 \end{array}}
	}{\{f\}}
	=
	\frac{1}{2}
	\dec{
		\ensuremath{\begin{array}{c}\begin{picture}(0,0)%
\includegraphics{pstex/Dumbbell-S-Sigma_g.pstex}%
\end{picture}%
\setlength{\unitlength}{3947sp}%
\begingroup\makeatletter\ifx\SetFigFont\undefined%
\gdef\SetFigFont#1#2#3#4#5{%
  \reset@font\fontsize{#1}{#2pt}%
  \fontfamily{#3}\fontseries{#4}\fontshape{#5}%
  \selectfont}%
\fi\endgroup%
\begin{picture}(320,932)(2178,-963)
\put(2247,-529){\makebox(0,0)[lb]{\smash{\SetFigFont{11}{13.2}{\rmdefault}{\mddefault}{\updefault}{\color[rgb]{0,0,0}$\DiagDot$}%
}}}
\put(2253,-225){\makebox(0,0)[lb]{\smash{\SetFigFont{11}{13.2}{\rmdefault}{\mddefault}{\updefault}{\color[rgb]{0,0,0}$\Sigma_g$}%
}}}
\put(2289,-859){\makebox(0,0)[lb]{\smash{\SetFigFont{11}{13.2}{\rmdefault}{\mddefault}{\updefault}{\color[rgb]{0,0,0}$S$}%
}}}
\end{picture}
 \end{array}} - \ensuremath{\begin{array}{c}\begin{picture}(0,0)%
\includegraphics{pstex/Padlock-Sigma_g.pstex}%
\end{picture}%
\setlength{\unitlength}{3947sp}%
\begingroup\makeatletter\ifx\SetFigFont\undefined%
\gdef\SetFigFont#1#2#3#4#5{%
  \reset@font\fontsize{#1}{#2pt}%
  \fontfamily{#3}\fontseries{#4}\fontshape{#5}%
  \selectfont}%
\fi\endgroup%
\begin{picture}(323,516)(1653,-510)
\put(1728,-386){\makebox(0,0)[lb]{\smash{{\SetFigFont{11}{13.2}{\rmdefault}{\mddefault}{\updefault}{\color[rgb]{0,0,0}$\Sigma_g$}%
}}}}
\put(1780,-114){\makebox(0,0)[lb]{\smash{{\SetFigFont{11}{13.2}{\rmdefault}{\mddefault}{\updefault}{\color[rgb]{0,0,0}$\DiagDot$}%
}}}}
\end{picture}%
 \end{array}}
	}{\{f\}}
	\]
%\resizebox{8.5cm}{!}{\includegraphics{figure1.epsi}}
\caption{The diagrammatic form of the QED
flow equation for vertices
of the Wilsonian effective action.}
\label{fig:Flow}
\ecf

The first term on the \lhs\ represents
the flow of all independent
Wilsonian effective action
vertex coefficient functions corresponding
to the set of fields, $\{f\}$.
Since the $\Lambda$-derivative
strikes just a vertex coefficient 
function---all fields
having been stripped off---we need not
write this as a partial derivative with
fields held constant [\cf~\eq{blocked}].
The term $\sum_{\phi \epsilon \{f\}} \gamma^{(\phi)}$
explicitly takes account of the anomalous
dimensions of the fields which suffer
field strength renormalization.
The field $\phi$ belongs to the set of fields 
$\{f\}$ and
the notation $\gamma^{(\phi)}$
just stands for the anomalous dimension of
the field $\phi$ (which, we recall, is zero for the 
gauge field, as a consequence of the manifest gauge
invariance).

The lobes on the \rhs\ of the flow equation
are vertex coefficient functions of
$S$ and $\Sigma_g$. These lobes are joined together
by the ERG kernels,
\DummyKernel, which are covariantized, as appropriate.
In QED it is necessary to covariantize only the kernels of the
fermions and their PV partners, meaning that these kernels
can be decorated by gauge fields.
The rule for decorating the diagrams on
the \rhs\ is simple: the set of fields, $\{f\}$, are distributed in
all allowed, independent ways between the component objects of each diagram.
For the details, the reader is referred to~\cite{qed}.

The understanding and efficient application of the diagrammatic flow equation
has been tremendously enhanced through a  diagrammatic calculus, proposed in~\cite{aprop}, refined in~\cite{mgierg1,mgierg2,primer,thesis,RG2005,mgiuc,evalues,qed,scalar2} and completed in~\cite{univ}. The central ingredient of this calculus is the `effective propagator relation'. The recent understanding of this relation~\cite{univ,NonRenorm} is as follows. Starting with the kernels, $\dd$, introduce the integrated kernels, $\Delta$, such that
\[
	-\totalflow \Delta \equiv \dd.
\]
The integrated kernels are what we refer to as the effective propagators.
Next, define a set of two-point vertices, $\itp$, that are essentially the inverses of the effective propagators.
Indeed, in the fermion and PV sectors, these vertices are precisely the inverses
of the corresponding effective propagators, but in the gauge sector things are more
subtle. As the name suggests, effective propagators are somewhat like usual propagators. In the fermion and PV sectors, they can be taken to be precisely UV regularized propagators~\cite{qed}. 
In the gauge sector, however, it is clear that we cannot interpret the integrated kernel simply as a regularized propagator, since we have not fixed the gauge and so cannot define a propagator in the usual sense! Nevertheless, there is nothing to stop us from defining ERG kernels and integrating them \wrt\ the effective scale. Now, when we come to contract the gauge sector effective propagator into the $\left( \itp \right)_{\mu \nu}(p)$ vertex, we should get the identity \emph{plus a remainder term}, where this remainder is forced upon us by gauge invariance. (This is all explained more fully in~\cite{aprop,qed,univ}.) Specifically, the gauge sector effective propagator is the inverse of the appropriate two-point vertex, in the transverse space:
\be
	\left( \itp \right)_{\mu \nu}(p) \Delta^{AA}(p) = \delta_{\mu \nu} - \frac{p_\mu p_\nu}{p^2}.
\label{eq:EPR-AA}
\ee
To give a specific example, let us introduce the UV cutoff function, $c(p)$, which satisfies $c(0) = 1$ and dies off sufficiently fast as $p^2/\Lambda^2 \rightarrow \infty$. We could now choose to identify
$\left( \itp \right)_{\mu \nu}(p)$ with the regularized classical two-point vertex, $c^{-1}(p)\Box_{\mu \nu}(p)$
and take $\Delta^{AA}(p) = c(p)/p^2$, which clearly satisfies~\eq{EPR-AA} (we have defined $\Box_{\mu \nu}(p) \equiv p^2 \delta_{\mu \nu} - p_\mu p_\nu$).

The reason that the effective propagator relation is so useful is that it allows the simplification of
a certain class of diagrams, which then allows the cancellation of other diagrams generated in
a typical calculation (see, particularly, \cite{mgiuc}). In the fermion and regulator sectors, this is the end of the story. In the gauge sector, we are left over with the remainders. However, it turns out that these can be processed diagrammatically, using the Ward identities~\cite{qed,mgiuc}, and the whole procedure of applying the effective propagator relation and cancelling terms can be iterated. As we shall see shortly, one result of these cancellations is that the $\beta$-function possesses no explicit dependence on either the seed action or the details of the covariantization of the cutoff. Looking at \fig{Flow}, this is really rather remarkable. Given all these cancellations, what is it that the $\beta$-function depends on? The answer is simply the exact $n$-point vertices, with all instances of $\itp$ having been extracted, joined together by effective propagators. Indeed, since instances of $\itp$ are removed via application of the effective propagator relation, it is useful to define reduced vertices according to:
\be
\label{eq:reducedWEA}
	\dec{\ensuremath{\begin{array}{c}\begin{picture}(0,0)%
\epsfig{file=pstex/ReducedWEA.pstex}%
\end{picture}%
\setlength{\unitlength}{3947sp}%
\begingroup\makeatletter\ifx\SetFigFont\undefined%
\gdef\SetFigFont#1#2#3#4#5{%
  \reset@font\fontsize{#1}{#2pt}%
  \fontfamily{#3}\fontseries{#4}\fontshape{#5}%
  \selectfont}%
\fi\endgroup%
\begin{picture}(358,358)(2279,-558)
\put(2336,-447){\makebox(0,0)[lb]{\smash{{\SetFigFont{11}{13.2}{\rmdefault}{\mddefault}{\updefault}{\color[rgb]{0,0,0}$\nothing S^{\mathrm{R}}$}%
}}}}
\end{picture}%
 \end{array}}}{\{f\}} \equiv
		\ds
		\dec{\ensuremath{\begin{array}{c}\begin{picture}(0,0)%
\includegraphics{pstex/WEA.pstex}%
\end{picture}%
\setlength{\unitlength}{3947sp}%
\begingroup\makeatletter\ifx\SetFigFont\undefined%
\gdef\SetFigFont#1#2#3#4#5{%
  \reset@font\fontsize{#1}{#2pt}%
  \fontfamily{#3}\fontseries{#4}\fontshape{#5}%
  \selectfont}%
\fi\endgroup%
\begin{picture}(358,358)(2279,-558)
\put(2392,-441){\makebox(0,0)[lb]{\smash{{\SetFigFont{11}{13.2}{\rmdefault}{\mddefault}{\updefault}{\color[rgb]{0,0,0}$\nothing S$}%
}}}}
\end{picture}%
 \end{array}} - \frac{1}{g^2} \ensuremath{\begin{array}{c}\begin{picture}(0,0)%
\includegraphics{pstex/ctp.pstex}%
\end{picture}%
\setlength{\unitlength}{3947sp}%
\begingroup\makeatletter\ifx\SetFigFont\undefined%
\gdef\SetFigFont#1#2#3#4#5{%
  \reset@font\fontsize{#1}{#2pt}%
  \fontfamily{#3}\fontseries{#4}\fontshape{#5}%
  \selectfont}%
\fi\endgroup%
\begin{picture}(400,398)(2370,-1145)
\put(2418,-1008){\makebox(0,0)[lb]{\smash{{\SetFigFont{11}{13.2}{\rmdefault}{\mddefault}{\updefault}{\color[rgb]{0,0,0}$\itp$}%
}}}}
\end{picture}%
 \end{array}} \delta_{2,n_f}}{\{f\}},
\ee
where $n_f$ is the number of fields in the set $\{f\}$.
If we chose to identify the $\itp$ vertices with the canonical classical, two-point vertices, then the reduced vertices are simply the vertices of the interaction part of the Wilsonian effective action.

Our aim, now, is to use the diagrammatic form of the flow
equation to compute the flow of a special combination of
diagrams. Following~\cite{mgiuc,univ,evalues,NonRenorm},
consider the following diagrammatic expression, which basically
constitutes all connected diagrams, possessing two external gauge fields, and
built from \emph{exact} $n$-point vertices:
\be
	\DV_{\mu \nu}(p) \equiv \sum_{s=0}^{\infty} \sum_{j=1}^{s+1} \norm_{s,j} g^{2s}
		\dec{\dec{\ensuremath{\begin{array}{c}\begin{picture}(0,0)%
\epsfig{file=pstex/ReducedWEA.pstex}%
\end{picture}%
\setlength{\unitlength}{3947sp}%
\begingroup\makeatletter\ifx\SetFigFont\undefined%
\gdef\SetFigFont#1#2#3#4#5{%
  \reset@font\fontsize{#1}{#2pt}%
  \fontfamily{#3}\fontseries{#4}\fontshape{#5}%
  \selectfont}%
\fi\endgroup%
\begin{picture}(358,358)(2279,-558)
\put(2336,-447){\makebox(0,0)[lb]{\smash{{\SetFigFont{11}{13.2}{\rmdefault}{\mddefault}{\updefault}{\color[rgb]{0,0,0}$\nothing S^{\mathrm{R}}$}%
}}}}
\end{picture}%
 \end{array}}}{j}}{\Delta^s A_\mu(p) A_\nu(-p)}
\label{eq:D}
\ee
where, for non-negative integers $a$ and $b$,
\be
\label{eq:norm}
	\norm_{a,b} \equiv \frac{(-1)^{b+1}}{a!b!} \left(\frac{1}{2}\right)^{a}.
\ee
We understand the notation of~\eq{D} as follows. The \rhs\ stands for all
independent, connected diagrams which can be created from $j$ reduced Wilsonian
effective action vertices, $s$ internal lines (\ie\ effective propagators)
and which are decorated by two external gauge fields, carrying momenta $p$ and $-p$ into the
vertex.
(It is the constraint of connectedness which restricts the sum over $j$.)
The combinatorics for generating fully fleshed out diagrams is simple
and intuitive and is described in~\cite{NonRenorm} (see also~\cite{RG2005,mgiuc}).
To gain a feeling for the structure of~\eq{D}, the first few terms represented by
the \rhs\ are shown in \fig{terms}.
\bcf[h]
	\[
	\DV_{\mu \nu}(p) = \ensuremath{\begin{array}{c}\begin{picture}(0,0)%
\epsfig{file=pstex/ReducedWEA-2.pstex}%
\end{picture}%
\setlength{\unitlength}{3947sp}%
\begingroup\makeatletter\ifx\SetFigFont\undefined%
\gdef\SetFigFont#1#2#3#4#5{%
  \reset@font\fontsize{#1}{#2pt}%
  \fontfamily{#3}\fontseries{#4}\fontshape{#5}%
  \selectfont}%
\fi\endgroup%
\begin{picture}(358,828)(1629,-796)
\put(1686,-451){\makebox(0,0)[lb]{\smash{{\SetFigFont{11}{13.2}{\rmdefault}{\mddefault}{\updefault}{\color[rgb]{0,0,0}$\nothing S^{\mathrm{R}}$}%
}}}}
\end{picture}%
 \end{array}} + \frac{1}{2} \ensuremath{\begin{array}{c}\begin{picture}(0,0)%
\epsfig{file=pstex/Padlock-2.pstex}%
\end{picture}%
\setlength{\unitlength}{3947sp}%
\begingroup\makeatletter\ifx\SetFigFont\undefined%
\gdef\SetFigFont#1#2#3#4#5{%
  \reset@font\fontsize{#1}{#2pt}%
  \fontfamily{#3}\fontseries{#4}\fontshape{#5}%
  \selectfont}%
\fi\endgroup%
\begin{picture}(629,630)(1493,-658)
\put(1754,-447){\makebox(0,0)[lb]{\smash{{\SetFigFont{11}{13.2}{\rmdefault}{\mddefault}{\updefault}{\color[rgb]{0,0,0}$\nothing S$}%
}}}}
\end{picture}%
 \end{array}} -\frac{1}{2}\ensuremath{\begin{array}{c}\begin{picture}(0,0)%
\epsfig{file=pstex/Thpt-Thpt.pstex}%
\end{picture}%
\setlength{\unitlength}{3947sp}%
\begingroup\makeatletter\ifx\SetFigFont\undefined%
\gdef\SetFigFont#1#2#3#4#5{%
  \reset@font\fontsize{#1}{#2pt}%
  \fontfamily{#3}\fontseries{#4}\fontshape{#5}%
  \selectfont}%
\fi\endgroup%
\begin{picture}(360,1387)(1631,-799)
\put(1754,-447){\makebox(0,0)[lb]{\smash{{\SetFigFont{11}{13.2}{\rmdefault}{\mddefault}{\updefault}{\color[rgb]{0,0,0}$\nothing S$}%
}}}}
\put(1756,109){\makebox(0,0)[lb]{\smash{{\SetFigFont{11}{13.2}{\rmdefault}{\mddefault}{\updefault}{\color[rgb]{0,0,0}$\nothing S$}%
}}}}
\end{picture}%
 \end{array}}
	+\cdots
	\]
\caption{A selection of the terms contributing to $\DV_{\mu \nu}(p)$ (Lorentz indices and momenta on the external lines have been suppressed). The flavours of the internal fields are
essentially summed over; for the precise statement of the Feynman rules, see~\cite{qed}.
Since reduction of the vertices only affects two-point vertices, we have removed the superscript `R' from all vertices with more than two legs.}
\label{fig:terms}
\ecf

Note that the diagrams of \fig{terms} have certain similarities to standard Feynman diagrams;
indeed, were we to shrink the lobes to points, they would look the same. However, despite
this similarity, our diagrams are related to the ERG flows of vertices of exact Wilsonian effective action
and not (directly) to perturbative scattering amplitudes. Whilst physics can most certainly be extracted from the Wilsonian effective action vertices, in the current case this must be done in a manifestly gauge invariant way~\cite{evalues}.

Defining $\OPI_{\mu \nu}$ to be the 1PI part of $\DV_{\mu \nu}$, with
\[
	\DV_{\mu \nu}(p) = \frac{\OPI_{\mu \nu}(p)}{1 + \Delta(p) \OPI_{\mu \nu}(p)},
\]
it is straightforward, but somewhat tedious, to use the diagrammatic calculus
to demonstrate that:
\be
	\frac{2 \beta}{g^3} \Box_{\mu \nu}(p)  + \order{p^4} = \totalflow \OPI_{\mu \nu}(p),
\label{eq:beta}
\ee
where
\be
	\beta \equiv \Lambda \der{g}{\Lambda}.
\label{eq:beta-defn}
\ee
(An explicit demonstration of many of the steps pertinent to this calculation are shown in the simpler
case of scalar field theory in~\cite{NonRenorm}. See also~\cite{mgiuc}.) Note that, as claimed earlier, the $\beta$-function has
been written in a form where there is no explicit dependence on either the seed action or the details of the
covariantization of the cutoff.

At this stage, we would do well to pause and carefully assess what kind of nonperturbative conclusions
we can reliably draw from~\eq{beta}.  The first point to make is that the entire diagrammatic approach relies on a weak field expansion, which has its drawbacks. For example, searches for nonperturbative fixed points, using such a scheme, rely on truncating the infinite tower of coupled equations for the exact $n$-point vertices, and this is known to give bad results~\cite{TRM-Truncations} (but see also~\cite{Aoki:1998um}). However, the situation is much better in the current case, as we now argue. First,  we never perform any truncations and will instead draw conclusions from general properties of the full function $\OPI_{\mu \nu}(p)$, which we emphasise depends on the \emph{exact} $n$-point vertices, no perturbative expansions having been performed. Secondly, we can always consider this function in the weak coupling regime. This does not mean to say that we wish to do perturbation theory, throwing away all nonperturbative contributions. Rather, we simply want to consider $\OPI_{\mu \nu}(p)$ in a regime
where its diagrammatic expansion can be ordered with a small parameter and could, at least in principle, be exactly resummed since we have not thrown any contributions away.

Thus, our understanding of $\OPI_{\mu \nu}(p)$ is as follows. Formally, it is given by its diagrammatic expansion~\eq{D} at all scales. More rigourously, this diagrammatic expansion should be evaluated in
the regime where the coupling is small, and resummed.

With these points in mind, we will now show that there cannot, in fact, be nonperturbative contributions
to~\eq{beta}, implying that the perturbative expansion of the $\beta$-function can actually be resummed, by itself. To do this, we re-express~\eq{beta} as:
\be
	\frac{2 \beta}{g^3} + \order{p^2} = \frac{\flow \OPI'(p)}{1-g^3 /2 \partial_g \OPI'(p)},
\label{eq:beta-partial}
\ee
where $\OPI'(p) \Box_{\mu \nu} (p) \equiv \OPI_{\mu \nu}(p)$ and the partial derivative \wrt\ $\Lambda$ is
performed at constant $g$. We now make the following observation: loop integrals in the diagrams comprising $\OPI'(p)$ can acquire factors of $\ln p^2 / \Lambda^2$, arising from IR divergences in the limit $p \rightarrow 0$. This is clear from analysing \eg\ the third diagram in \fig{terms}, for which the component which goes like 
\[
	\FourInt{k} \frac{1}{\ksl} \frac{1}{\psl + \ksl}
\]
in the IR (see~\cite{qed} for the details) produces the desired behaviour after we act on the full diagram  with the $\Lambda$-derivative and pull out $\Box_{\mu\nu}(p)$. 
[On dimensional grounds, we see that the integrand $\sim p^2/k^4$, at $\order{p^2}$.]
It is important to note that the apparent UV divergence in
$\ln p^2/\Lambda^2$ is an artefact of us having Taylor expanded vertices and cutoff functions
in the external momentum (as it must be: by construction, everything is UV regularized). Indeed, as mentioned earlier, all vertices have a derivative expansion, as
do the cutoff functions. The only non-polynomial dependence of $\OPI'(p)$ on the external momentum is
generated by certain loop integrals in the IR.

Furthermore,
whilst individual diagrams contributing to $\OPI'(p) +\order{p^2}$ can diverge as a logarithm to a power
(which is at most equal to the number of loops) as $p \rightarrow 0$, all divergences must cancel out between the numerator and denominator of~\eq{beta-partial} since the $\order{p^0}$ contribution to the \lhs\ of~\eq{beta-partial} has no additional, non-polynomial dependence on $p$. 
Consequently, for functions $F_1$, $F_2$ and $G$, it must be that we can write:
\be
	\frac{2 \beta}{g^3} + \order{p^2} = \frac{F_1(g^2) G(g^2, \ln p^2/\Lambda^2)}{F_2(g^2) G(g^2, \ln p^2/\Lambda^2)} = \frac{F_1(g^2)}{F_2(g^2)}.
\label{eq:beta-form}
\ee

So, to begin with, let us consider perturbative contributions in~\eq{beta-partial}. Let us suppose that, 
at order $g^{2n}$, the strongest IR divergence carried by $\OPI'(p)$ goes like
\be
	g^{2n} \ln^m p^2 /\Lambda^2.
\label{eq:div-pert}
\ee
In the numerator, the $\Lambda$-derivative reduces this divergence to one of the form
\be
	g^{2n} \ln^{m-1} p^2 /\Lambda^2
\label{eq:num-div}
\ee
whereas, in the denominator, a contributions of the form
\be
	g^{2(n+1)} \ln^m p^2 /\Lambda^2
\label{eq:den-div}
\ee
is produced.
Thus, we have found that terms of the form~\eq{div-pert} provide a divergent contribution to the denominator which
does not seem to exist in the numerator. Of course, there is no real problem here: all we need to do is consider diagrams with an extra loop. In such diagrams there are
contributions of the form~\eq{div-pert} but with $n \rightarrow n+1$ and $m \rightarrow m+1$. Terms like this in the numerator are, after differentiation \wrt\ $\Lambda$, of precisely the right form to cancel denominator contributions of the type~\eq{den-div}. This is explicitly borne out in perturbative calculations~\cite{mgierg2,scalar2}.

But now consider a contribution of the type
\be
	g^{2n} \e{-a/g^2} \ln^m p^2 /\Lambda^2,
\label{eq:div-nonpert}
\ee
where again we assume that, for our choice of $n$, there is no stronger IR divergence. In the numerator this contributes terms of the form
\be
	g^{2n} \e{-a/g^2} \ln^{m-1} p^2 /\Lambda^2
\label{eq:num-np-div}
\ee
and in the denominator it yields terms of the form
\be
	g^{2n} \e{-a/g^2} \ln^m p^2 /\Lambda^2 + \ldots,
\label{eq:den-np-div}
\ee
where the ellipsis denotes terms higher order in $g^2$. Crucially, 
\eqs{num-np-div}{den-np-div} are \emph{the same order} in $g^2$.
Since, by assumption, there are no terms in $\OPI'(p)$ which
are of order $g^{2n}\e{-a/g^2}$ but which have a stronger IR divergence than~\eq{div-nonpert},
there is no way that the denominator contribution~\eq{den-np-div}
can ever be cancelled. From~\eq{beta-form}, we therefore conclude
that
terms of the
type~\eq{div-nonpert} must be absent from~\eq{beta-form}, unless $m=0$.
But it is easy to see that $m=0$ terms can appear only in $G(g^2,\ln p^2/\Lambda^2)$
and not in $F_1(g^2)$ or $F_2(g^2)$: for if this condition is violated, then
we necessarily produce contributions of the form~\eq{div-nonpert}, when
we expand out $F_1(g^2) G(g^2, \ln p^2/\Lambda^2)$. In conclusion, 
the only contributions to the $\beta$-function 
of the form~\eq{div-nonpert} that are allowed---namely
those with $m=0$---cancel out!

It is now straightforward to generalize this argument to show that only
the perturbative contributions to the $\beta$-function survive. First, we
note that the above argument is not affected if we consider terms
which include $\e{-b/g^4}$, $\e{-c/g^6}$ \etc, or products of such terms.
Secondly, we can allow additional functions of $g$ to come along
for the ride, so long as they do not spoil the requirement that the ERG
trajectory sinks into the Gaussian fixed point as $\Lambda \rightarrow 0$.

Thus, we have demonstrated that the $\beta$-function is free of nonperturbative power corrections and, therefore, must be resummable, at least in the massless theory. In the presence of a fermion mass, $\mu$, there would be no good reason to exclude surviving contributions to the $\beta$-function which
go like
\[
	\frac{\mu^2}{\Lambda^2} \e{-a/g^2},
\]
since the mass now regularizes terms which previously diverged as $p \rightarrow 0$.
Note, though, that as emphasised in~\cite{BenekeReview}, the presence of exponentially small terms does not, by itself, necessarily imply that perturbation theory cannot be resummed (though it is suggestive). In other words, it is at least possible that a (non-analytic) function comprises a resummable perturbative series plus additional exponentially small terms. Of course, were this scenario to be realized in the massive case, one would
certainly have to provide an argument as to why the perturbative series was free of renormalons. 

Returning to the massless theory, 
the resummability of the $\beta$-function is valid in the infinite number of schemes implicit to our approach; these different schemes corresponding to all the legal choices of the seed action and covariantization of the cutoff. There is no reason to expect that this conclusion is true for unrelated schemes, such as $\overline{MS}$.
It is important to add that no  expression as neat as~\eq{beta} exists for the flow of the other couplings
or for the  anomalous dimension, $\gamma$ (see~\cite{qed,mgiuc,NonRenorm} for the tools necessary to compute these expressions). Consequently, there is no way to argue that the
perturbative series for these functions, also, are resummable. Indeed, we expect precisely the
opposite to be true, since we do not expect self-similar trajectories to exist within the critical surface
of the Gaussian fixed point. Nevertheless, it would doubtless be interesting to compute the $\beta$-function to some high order in perturbation theory and resum it, not least from the point of view
of understanding the fate of the Landau pole in the massless theory. Perhaps more interesting still
would be to try to get some handle on what happens in the massive case, particularly given
the work already done on ERG flows in QED~\cite{Gies-QED}.

Finally, we should note that one can attempt to repeat the above analysis for other field theories.
In QCD, the expression for the $\beta$-function possesses additional terms, which can spoil the above arguments (the basic structure is apparent at the perturbative level~\cite{mgiuc}). Nevertheless, for this to happen, there must be some delicate relationships
between the various terms contributing to the $\beta$-function, which merits further 
investigation~\cite{WIP}. In scalar field theory the $\beta$-function has a sufficiently complicated form to spoil the above arguments, as expected~\cite{TRM-ApproxSolns,TRM-Elements}. 
The most interesting
case to look at, as mentioned earlier, is the Wess-Zumino model, where the above resummability argument seems to apply to the anomalous dimension (equivalently the $\beta$-function if we induce a flow of the superpotential by using the field strength renormalization to rescale the field). This is exciting because, unlike in QED, there are arguments to suggest that resummability of the anomalous
dimension implies resummability of the entire perturbative expansion of the theory.
If true, this would be highly suggestive of the existence of a UV fixed point~\cite{WIP}.

\begin{acknowledgments}
 I acknowledge IRCSET for financial support. I would like to thank Denjoe O'Connor
 for useful discussions.
\end{acknowledgments}

\bibliography{Resum2}

\end{document}